\documentclass[12pt]{article}
\usepackage{group21}
\title{ Resonances from Higher Order S-Matrix Poles with Exponential Decay}
\author{Christoph P\"{u}ntmann\thanks{talk presented at the Group21 
Conference in Goslar} \ and\ Paul Patuleanu} 
\address{Center for Particle Physics, ~~Department of Physics,\\
	The University of Texas at Austin, ~~Austin, Texas~78712-1081}
\begin{document}
\maketitle
\begin{abstract}
In analogy to Gamow vectors describing resonance states from first
order S-matrix poles, one can define Gamow vectors from higher order
poles of the S-matrix. With these vectors we are going to discuss 
a density operator that describes exponentially decaying resonances from
higher order poles.  
\end{abstract}

\vspace{12pt}
Singularities of the analytically continued S-matrix have been used to
describe resonances that decay exponentially~\cite{newton}. But these
states could not be described in ordinary Hilbert space quantum
mechanics which doesn't provide the elements to describe the purely
exponentially decaying vectors~\cite{fonda}. Resonances from 
higher order poles, in particular double poles, have also
been mentioned~\cite{goldberger,goldhaber}, 
but were always associated with an additional polynomial time
dependence~\cite{goldberger}. However, operators in the
form of finite dimensional matrices
consisting of non-diagonalizable Jordan blocks
have been discussed in connection with resonances numerous times in
the past~\cite{mondragon}
%~\cite{mondragon,stodolsky,lukierski,brandas,antoniou-tasaki}. 

With the introduction of the rigged Hilbert space~\cite{gelfand} it
was possible to describe resonances~\cite{bohm,BG,bohm-gadella}, 
with the features usually attributed to resonances, i.e.
exponential decay law and Breit-Wigner energy 
distribution. They are constructed from the
first order poles of the analytically continued S-matrix on the
second Riemann sheet of the complex energy plane, which come
in pairs above and below the real axis. The pole at
$z_R=E_R-i\Gamma/2$ corresponds to the decaying state defined for 
times $t\geq0$, and the pole at $z^*_R=E_R+i\Gamma/2$ to the respective 
growing state for $t<0$. 
The Gamow vectors are generalized eigenvectors of a self-adjoint
Hamiltonian $H$, whose adjoint $H^\times$ in the rigged Hilbert space 
is an extension of the adjoint $H^\dagger=H$ in the Hilbert space,
with complex eigenvalues $E_R\pm i\Gamma/2$ (energy and lifetime).
These vectors form a complex basis system expansion, 
in which the Hamiltonian can be represented by a diagonal matrix with
complex energies on the diagonal. 
The time evolution is given by a semigroup operator~\cite{kielanowski}
which time translates the decaying state vectors for $t\geq0$, and the
growing state vectors for $t<0$.
%The time evolution operator is given by the semigroup
%operator~\cite{kielanowski}
%$e^{-iH^\times t}$ which for times $t<0$ time-translates the growing
%state $|z^+_R\rangle$ according to
%\begin{equation}
%	e^{-iH^\times t}|z^+_R\rangle
%	=e^{-iz^*_Rt}|z^+_R\rangle\,,\;\;t<0
%\end{equation}
%and for times $t>0$ time-translates the decaying state $|z^-_R\rangle$
%according to
%\begin{equation}
%	e^{-iH^\times t}|z^-_R\rangle
%	=e^{-iz_Rt}|z^-_R\rangle\,,\;\;t>0\,.
%\end{equation}

\vspace{12pt}
The mathematical procedure by which these Gamow vectors were
introduced suggests a straightforward generalization to higher order
Gamow vectors which are derived from higher order
poles of the S-matrix~\cite{antoniou-gadella}. It 
can be shown that the $r$-th order pole of a unitary S-matrix leads to
$r$ generalized vectors of order $k=0,~1,~\dots,~r-1$,~\cite{JMP}
\begin{equation}\nonumber
	|z^-_R\rangle^{(0)}\,,\;|z^-_R\rangle^{(1)}\,,\;\cdots,
	|z^-_R\rangle^{(k)}\,,\cdots,|z^-_R\rangle^{(r-1)}\;.
\end{equation}
In the same way one derives an analogous set of $r$ generalized vectors
for the S-matrix pole at $z^*_R$ associated with the growing state.
The higher order Gamow vectors form a complex basis vector expansion that
spans an $r$-dimensional subspace ${\cal M}_{z_R}$ of the RHS.
They are Jordan vectors~\cite{baumgartel} of degree $k+1$, 
and they are generalized eigenvectors~\cite{lancaster} 
of a self-adjoint Hamiltonian with the complex eigenvalue 
$z_R=E_R-i\Gamma/2$ such that
\begin{eqnarray}
	H^\times |z^-_R\rangle^{(0)} &=& z_R\, |z^-_R\rangle^{(0)} \nonumber \\
	H^\times |z^-_R\rangle^{(k)}&=&z_R|z^-_R\rangle^{(k)}
	+k|z^-_R\rangle^{(k-1)};\;\;k=1,\dots,r-1\,.\nonumber
\end{eqnarray}
This means that $H^\times|_{{\cal M}_{z_R}}$ is represented by a finite
dimensional Jordan block matrix of degree $r$.

The time evolution of the higher order Gamow vectors has a polynomial
time dependence besides the exponential:
\begin{equation}\nonumber
	e^{-iH^\times t}|z^-_R\rangle^{(k)}=e^{-iz_Rt}
	\sum^k_{\nu=0}\left(\begin{array}{c}k\\\nu\end{array}\right)\,
	(-it)^\nu\,|z^-_R\rangle^{(k-\nu)}\hspace{.2in}t>0\,.
\end{equation}
The semigroup operator $e^{-iH^\times t}$ transforms between different
$|z^-_R\rangle^{(k)}$ that belong to the same pole of order $r$ at
$z_R$, but it does not transform out of ${\cal M}_{z_R}$. A
higher order Gamow vector of degree $k+1$ is transformed into a
superposition of higher order Gamow vectors of the same and all lower
degrees. The time evolution of the higher order Gamow vectors leads,
as in the case of the ordinary Gamow vectors, to an intrinsic
microphysical arrow of time~\cite{kielanowski}.

The label $k$ of the higher order Gamow vectors is not a quantum
number in the usual sense. Basis vectors are usually labeled by
quantum numbers associated with eigenvalues of a complete system of
commuting observables (\cite{bohm}, chap.~IV), but there is no
physical observable that the label $k$ is connected to. Therefore, the
different $|z^-_R\rangle^{(k)}$ in the subspace ${\cal M}_{z_R}$
do not have a separate physical meaning.

\vspace{12pt}
In analogy to von Neumann's description of physical states by dyadic
products of state vectors, Gamow states have been described by dyadic
products of the ordinary Gamow vectors~\cite{bohm}.
Examples of these states with their exponential decay law and their
Breit-Wigner energy distribution have been observed in abundance as
resonances and decaying states. 
Theoretically, there is no reason to exclude quasistationary states
from higher order poles. One argument made against their existence 
was the polynomial time dependence that was vaguely associated with 
them and which has not been observed.

For a microphysical decaying state associated with an $(n+1)$-st order
S-matrix pole, the structure of the complex basis vector expansion~\cite{JMP}
(pole term of the S-matrix element) suggests as a form of a higher
order Gamow density operator:
\begin{equation}\label{k}
	W^{(n)}=\sum_{k=0}^n\left(\begin{array}{c}n\\k\end{array}\right)
	|z^-_R\rangle^{(k)}\;^{(n-k)}\langle^-z_R|\,.
\end{equation}
This microphysical state is a mixture of non-reducible components. In
spite of the fact that the higher order Gamow vectors have an
additional polynomial time dependence, this microphysical state obeys
a purely exponential decay law~\cite{JMP},
\begin{equation}\nonumber
	W^{(n)}(t)=e^{-\Gamma t} \, W^{(n)}(0)\,; \hspace{.2in} t\geq 0\,.
\end{equation}
However, they do not describe resonances with a simple 
Breit-Wigner energy distribution. Instead their energy distribution is
a sum of the Breit-Wigner energy distribution and its derivatives up
to order $n=r-1$.%~\cite{GR} 
It can be shown that these density operators are the
only operators that can be constructed from the higher order Gamow
vectors (describing resonances from higher order S-matrix
poles) that lead to an exponential decay law~\cite{uniqueness}. 
In the zeroth order case one trivially deals with a pure state.
However, for higher order states, ``pure'' has probably no meaning,
since $k$ is not a quantum number connected with a physical
observable. 

%The density operator~(\ref{k}) is a non-reducible ``mixture''
%~\cite{prigogine} 
%which describes
%resonances from higher order poles with exponential decay. However,
%they do not describe resonances with a simple 
%Breit-Wigner energy distribution. Instead
%their energy distribution is
%a sum of the Breit-Wigner energy distribution and its derivatives up
%to order $n=r-1$.~\cite{GR} 

%\vspace{12pt}
At presence there is little empirical evidence for the existence of
these higher order pole states in nature. Our results suggest that the
empirical objection to the existence of higher order poles of the
S-matrix does not rule out the possibility of exponentially decaying
states constructed from higher order Gamow vectors.

\vspace{12pt}
We are very grateful to the organizers of the conference Group21 in Goslar, 
especially to H.~D. Doebner and W.~Scherer. Also, we wish to 
acknowledge A.~Bohm, M.~Gadella, and M.~Loewe for their collaboration
on this subject.

\end{document}